# Robustness of the pyrochlore structure in rare-earth $A_2Ir_2O_7$ iridates and pressure-induced structural transformation in $IrO_2$

Daniel Staško[1,*], Kristina Vlášková[1], Andrej Kancko[1], Daniel M. Többens[2], Dominik Daisenberger[3], Gaston Garbarino[4], Ross Harvey Colman[1], Milan Klicpera[1]

[1]Charles University, Faculty of Mathematics and Physics, Department of Condensed Matter Physics, Ke Karlovu 5, 12116 Prague 2, Czech Republic

[2]Helmholtz-Zentrum Berlin, Department Structure and Dynamics of Energy Materials, Albert-Einstein-Straße 15, 12489 Berlin, Germany

[3]Diamond Light Source Ltd, Harwell Campus, Didcot OX11 0DE, UK

[4]ESRF, The European Synchrotron, 71 Avenue des Martyrs, CS40220, 38043 Grenoble Cedex 9, France

*Corresponding author: daniel.stasko@matfyz.cuni.cz

**Abstract:**

A comprehensive study of the structural properties of the heavily investigated rare-earth $A_2Ir_2O_7$ series under extreme conditions is presented. The series is covered by studying iridates with A = Pr, Sm, Dy-Lu. Temperature- and pressure-dependent synchrotron X-ray powder diffraction experiments reveal robustness of the pyrochlore structure throughout the rare-earth series, down to 4 K and up to 20 GPa. The thermal expansivity and pressure compressibility are determined, including Debye temperature, bulk modulus and Grüneisen parameter. Temperature and pressure evolution of the fractional coordinate of oxygen at 48$f$ Wyckoff position, the sole free atomic position parameter, is investigated and discussed regarding the antiferromagnetic ordering of the Ir magnetic moments. In addition, the pressure evolution of the crystal structure of an $IrO_2$ minority phase is followed. The tetragonal rutile-type structure is orthorhombically distorted at 15 GPa, and the orthorhombic structure is not fully stabilised up to 20 GPa.

1. Introduction:

Rare-earth iridium oxides $A_2Ir_2O_7$ have attracted considerable attention of the scientific community for their frequently complex magnetic and conductive properties. A variety of observed and predicted phenomena in these materials, including topological Mott insulator [1], axion insulator [2], Weyl semimetal [3,4], spin ice [5], spin liquid [6], or fragmented states [5,7], originate in concurrence of several mechanisms: (i) a strong spin-orbit coupling (SOC) related to rare-earth elements and mainly Ir, (ii) an intermediately strong Coulomb repulsion in heavy elements, (iii) exchange magnetic interactions, (iv) d-f exchange between Ir and A moments, (v) geometrical frustration of the crystal lattice, and (vi) crystal field effects.

In general, the strength of the spin-orbit interaction increases with the atomic number of the element, while the electronic orbitals become more extended, leading to weaker Coulomb repulsion. Mutual strengths of the SOC and Coulomb interaction (comparable in iridates) determine, inter alia, the conductive properties of the material [2]. Indeed, various electronic properties have been observed in

$A_2Ir_2O_7$ iridates, such as an unconventional anomalous Hall effect in $Pr_2Ir_2O_7$ [8], giant magnetoresistance in $Nd_2Ir_2O_7$ [9], and Weyl semimetal charge transport behaviour in $Y_2Ir_2O_7$ [10]. A metal-insulator transition (MIT) is found in most of the series (A = Nd-Lu), concomitant with an antiferromagnetic ordering of the Ir sublattice. A continuous change in the MIT/magnetic ordering temperature, $T_{Ir}$, appears to depend primarily on the A-site cation size [11,12]. Such a smooth behaviour, mostly irrespective of the particular A-site magnetic and electronic properties, has been attributed to the wide $t_{2g}$-block bandwidth of iridium resulting in a strong 5d-orbital overlap with oxygen ligand 3p-orbitals, and a strong sensitivity to structural changes (especially Ir-O bond length and Ir-O-Ir bond angle variation) associated with the pyrochlore lattice contraction [13]. At the same time, strong electron-phonon interactions have been reported for $Eu_2Ir_2O_7$ [14], revealing softening and line-shape anomalies of the Ir-O-Ir bond-bending vibration [15] and magnon mode softening at $T_{Ir}$.

The MIT, $T_{Ir}$, and generally the conducting properties of $A_2Ir_2O_7$ can be tuned by changing the structural parameters of the system. Three systematic approaches are available for studying perturbations in this system: (i) substitution of elements on the A-site [16,17,18], (ii) substitution on the Ir-site [19,20], or (iii) application of external pressure. External pressure represents a clean way to study lattice properties, avoiding the introduction of atomic disorder into the lattice by partial site substitution. The inequivalence of chemical and physical pressure was documented, e.g., in $Sm_2Ir_2O_7$ [21]. So far, the experiments under pressure were limited to the light-A pyrochlores. Applying pressure on $Nd_2Ir_2O_7$ causes a gradual decrease of $T_{Ir}$ (= 36 K at ambient pressure), suppression of MIT at 10 GPa, and further pressure application leads to an emergence of a new magnetic phase [22]. On the other hand, the transition temperature in $Eu_2Ir_2O_7$ ($T_{Ir}$ = 120 K) decreases only slightly by increasing the pressure up to 6 GPa, and a subtle increase is observed by further pressure application [23]. Simultaneously, the character of the metal changes from incoherent (high resistivity) to a more conventional one with increasing pressure. A non-monotonic development of $T_{Ir}$ in $Eu_2Ir_2O_7$ was further confirmed by susceptibility measurement (up to approximately 4.2 GPa) [24]. However, in contrast, $T_{Ir}$ slightly increased with applied pressure up to 2.2 GPa and decreased again with further pressure application. Heavy-A members of the series have remained mostly uninvestigated from the viewpoint of external pressure application, although the possibility to further tune the behaviour between the various predicted unconventional states using pressure is of particular interest [1,25].

Complementary structural and compressibility information is imperative to properly interpret both current and future pressure-dependent properties of these compounds. The pyrochlore structure of $A_2Ir_2O_7$ (space group *Fd-3m*, found in the majority of the $A_2B_2O_7$ oxides) is a perfectly ordered cubic structure with only two free parameters: the lattice parameter *a* and the fractional coordinate of oxygen at the 48f Wyckoff position $x_{48f}$. Related Ir-O bond lengths and Ir-O-Ir bond angles play a crucial role in interpreting the varying electronic and magnetic properties in $A_2Ir_2O_7$ materials [13,26]. Recently, anomalies in both the bond length and the bond angle developments were observed around $T_{Ir}$ in $Eu_2Ir_2O_7$ [25]. Details of the crystal structure of other $A_2Ir_2O_7$ members, not only at temperatures close to $T_{Ir}$, but throughout the low-temperature region, were generally missing. We report on the temperature, pressure and A dependence of the structural parameters of selected pyrochlore $A_2Ir_2O_7$ iridates (A = Pr, Sm, Dy-Lu), including the significantly understudied heavy-rare-earth members of the series, inclusive of newly synthesised $Tm_2Ir_2O_7$.

2. **Sample synthesis and experimental details:**

Polycrystalline $A_2Ir_2O_7$ samples (A = Pr, Sm, Dy-Lu) were synthesised employing a CsCl-flux method. Initial $A_2O_3$ and $IrO_2$ oxides (99.99% metal basis, Alfa Aesar) in a stoichiometric ratio were properly mixed with the CsCl flux (ratio $A_2O_3/IrO_2/CsCl$ = 1:2:50). The mixture was repeatedly reacted in a Pt crucible in air at 1073 K. A detailed description of the synthesis process is demonstrated on an example of $Er_2Ir_2O_7$ in our recent publication [27]. The quality of the reacted material was checked by X-ray diffraction (Bruker diffractometer, Cu Kα radiation) and scanning electron microscopy (TESCAN, Mira I LMH) with an energy-dispersive X-ray (EDX) analyser. The crystal structure of all synthesised samples was identified at room temperature as the cubic pyrochlore type (Fig. 1). Although the oxygen content in the sample cannot be reliably determined employing EDX analysis, the A:Ir ratio was checked. Small variations in off-stoichiometry (Ir deficiency) were observed in most samples (described in more detail below). Both methods revealed the presence of minority phases in the samples: $A_2O_3$, $IrO_2$, and Ir. These minority phases, that is, initial oxides and Ir from the dissociation of $IrO_2$, are generally present in synthesised samples as numerously reported for $A_2Ir_2O_7$ compounds [17,18,19,20,24].

The temperature and pressure developments of the structural parameters of individual $A_2Ir_2O_7$ members were investigated by means of powder diffraction experiments using X-ray synchrotron radiation. The temperature evolution of the diffraction patterns was measured employing the ID22 beamline [28] at the ESRF (European Synchrotron Radiation Facility) in Grenoble, France; and the KMC-2 beamline [29] at the Bessy II synchrotron facility in Berlin, Germany. The samples were loaded into 0.3 mm capillaries (ID22) and circular sample holder with a diameter of 10 mm and a thickness of 0.5 mm (KMC-2), respectively. Measurements of single high-statistics patterns using the X-rays' wavelength of 0.35427632 Å (ID22) and 1.1266 Å (KMC-2) were performed at selected temperatures covering an interval from 4 and 25 K to 295 K, with a step of 5 K in the vicinity of $T_{Ir}$ and a larger step of 10-15 K at other temperatures. The ID22 data at temperatures below 50 K were measured attenuating the high-intensity beam. High-pressure diffraction experiments were carried out using the ID15b beamline [30] at the ESRF and the I15 beamline [31] at the Diamond Light Source (Didcot, United Kingdom) with DACs (diamond anvil cell) employed to reach 20 GPa of pressure. Diamonds with a culet diameter of 400 μm and 600 μm were used at I15 and ID15b, respectively. Using X-rays' wavelength of 0.4102 Å (ID15b) and 0.3542 Å (I15), the pressure dependencies were measured at room temperature with a pressure step of approximately 0.25 GPa at low pressures and larger steps in higher pressure interval. Gas membrane DAC allowed in-situ pressure changes. Helium was used as the pressure-transmitting medium and the pressure was determined using ruby fluorescence [32]. 2D diffraction patterns measured at high pressures were processed (masked and integrated) using the Dioptas software [32]. All integrated data were refined using the Rietveld method with Topas academic [33]. The EOSfit GUI software [34] was employed for the equation-of-state analysis of the compressibility data.

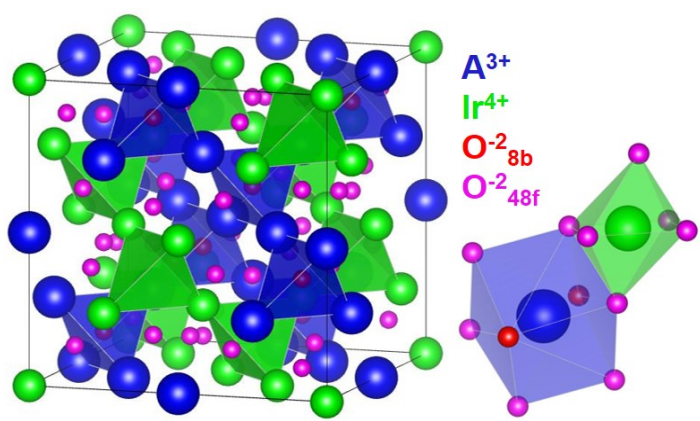

Fig. 1: Pyrochlore structure, cubic $Fd\text{-}3m$ space group, with the rare-earth $A^{3+}$ and $Ir^{4+}$ ions separately creating sublattices of corner-sharing tetrahedra. All oxygen anions at the 8b Wyckoff positions (red) are inside the $A^{3+}$ and $Ir^{4+}$ tetrahedra while the ones at the 48f position (pink) are outside. The $A^{3+}$ cations are surrounded by eight-coordinate oxygen cages and $Ir^{4+}$ by six-coordinate distorted-octahedral cages. In the case of $Ir^{4+}$, all six surrounding oxygen ions are at the 48f Wyckoff position, with the same Ir-O

bond lengths. The structural parameter $x_{48f}$ dictates the distortion of the octahedra (bond angles); perfectly symmetric octahedra exist only in a special case when $x_{48f} = 0.3125$ (which is not the case for the rare-earth iridates, where typically higher values of $x_{48f}$ are determined).

### 3. Orthorhombic distortion of the IrO$_2$ tetragonal structure induced by applied pressure

In addition to the details and evolution of the pyrochlore structure in the A$_2$Ir$_2$O$_7$ iridates (reported below), the experimental data provided information on a structural transition in the minority IrO$_2$ phase. As previously mentioned, IrO$_2$ is one of the minority phases commonly found in the A$_2$Ir$_2$O$_7$ powder samples [17,18,19,20,24]. Most samples revealed only weak or negligible signal related to the IrO$_2$ minority phase; the Tm$_2$Ir$_2$O$_7$ sample with approximately 2.7% IrO$_2$ phase in the content was an exception and allowed us to dependably study the stability of the IrO$_2$ crystal structure. Still, the intensity of the respective peaks was two orders of magnitude weaker than that of pyrochlore peaks. In fact, the IrO$_2$ peaks had only slightly higher intensity than the background signal, and they were often in close vicinity or overlapping with the pyrochlore peaks. Therefore, a quantitative analysis was difficult, especially in the case of the high-pressure data.

IrO$_2$ crystallises in the tetragonal rutile-type structure (space group $P4_2/mnm$) in ambient conditions [35,36]. We confirmed this crystal structure, refining our synchrotron data collected at temperatures down to 4 K and pressures up to 14 GPa. A standard contraction of the tetragonal lattice with ambient conditions parameters $a = 4.499(1)$ Å and $c = 3.153(1)$ Å (Ir occupies the 2$a$ Wyckoff position, O adopts the 4$d$ position) was followed. Upon reaching the pressure of 15 GPa, a considerable change of diffraction patterns was observed. The pronounced splitting of six out of eight IrO$_2$ peaks unambiguously identified in the diffraction patterns was revealed. See an example of the 211 peak splitting under applied pressure in Fig. 2b. With further increasing pressure, the splitting became more pronounced. A strong lattice distortion was followed up to 20 GPa and is expected to be even more significant with further increasing pressure. The remaining two peaks (out of eight) exhibit no splitting. Nevertheless, their evolution with pressure apparently changes at 15 GPa (see Fig. 2a with 110 reflection as an example), indicating a higher compressibility of the high-pressure phase. Additional peaks corresponding to the IrO$_2$ minority phase were either overlapping with peaks of majority phase or other phases, or had very low intensity, making their determination highly ambiguous. List of all tracked peaks and their positions is presented in Table S1.

In previous studies, a high-pressure phase of IrO$_2$ was synthesised. A combination of high-pressure and high-temperature conditions was used during the preparation process (pressure of 15 GPa and temperature of 2000 K [35] or 17 GPa and 1273 K [36]). However, this phase appeared to be unstable upon cooling, leading to its gradual decomposition [35] unless quenched [36]. The high-pressure high-temperature phase was determined as a cubic pyrite-type structure ($Pa\bar{3}$) [35,36]. A stark difference to our work is the additional heating process that was not present in our experiment, resulting in a very slow phase transformation at ambient temperature. In turn, the high-pressure phase of IrO$_2$ revealed by our measurements cannot be identified with the previously reported cubic structure.

The evolution of diffraction peaks suggests a distortion of the tetragonal structure induced by an applied pressure. Following the evolution of the observed peaks, an orthorhombic distortion was considered. That is, the tetragonal $a$-axis splits into orthorhombic $a$- and $b$-axis, which would have the same length at 14 GPa, and their difference will increase with further increasing pressure. The $c$-axis, on the other hand, remains unchanged (suggested by an evolution of peaks identified as 110 and 220 in tetragonal notation). Indeed, determining the lattice parameters from 14 peaks' positions measured at 20 GPa led to $a = 4.555(5)$ Å, $b = 4.263(5)$ Å, and $c = 3.137(4)$ Å. Note the small change of the $c$ parameter (less than 1%; connected with a standard lattice compression) comparing the orthorhombic and the tetragonal lattice, and the significant distortion of the tetragonal basal plane (1% increase in *a* and 5% decrease in *b* compared to the tetragonal *a* parameter).

Determination of the space group of the orthorhombic structure and especially the Wyckoff positions of individual ions is complicated, and generally impossible, without quantitative information on the diffraction intensity. Nevertheless, further information can still be obtained by inspecting the diffraction patterns and performing the symmetry analysis: Simulation of diffraction patterns using a model orthorhombic space group *Pmmm* revealed a number of diffraction peaks that were not reproduced in the measured data. Additional symmetry operations of the lattice are necessary to induce systematic absences and restrict the number of observable reflections. We performed the symmetry analysis calculating the orthorhombic translation-subgroups of the parent $P4_2/mnm$ space group. Simulating diffraction patterns using these subgroups (*Cmmm, Pnnm, Amm2, Cmm2, Pnn2, Pmn2$_1$, C222,* and *P2$_1$2$_1$2*) resulted in only two models that lead to diffraction patterns consistent with the experiment.

To distinguish between the two space groups, *Pnnm* and *Pnn2*, and unambiguously determine the high-pressure structure is impossible on the basis of our data. In the case that the high-pressure structure is described by the *Pnnm* space group, the new atomic Wyckoff positions of Ir and O would be 2*a* and 4*f* (where a free *z*-coordinate of the 4f site cannot be determined), respectively. Considering the *Pnn2* space group, the Ir adopts the 2*a* position and O occupies two 2*b* sites (again, free $z_{2b1}$ and $z_{2b2}$ coordinates of the $2b^1$ and $2b^2$ sites cannot be determined). A reliable determination of the space group and atomic positions in high-pressure $IrO_2$ phase structure is desirable and could be straightforwardly done investigating pure $IrO_2$ sample in a future pressure experiment. Nevertheless, at the moment, we tend to describe the high-pressure phase of $IrO_2$ by space group *Pnnm*; that is, the higher-symmetry space group. Indeed, previous reports on several binary oxides document a transformation from the rutile-type structure ($P4_2/mnm$) to the $CaCl_2$-type structure (*Pnnm*) at high pressures [37,38]. Simultaneously, a second pressure induced transformation to the α-$PbO_2$-type structure (*Pbcn*) was reported for many analogues at higher pressure [39,40]. First-principle calculations study [41] revealed that both $CaCl_2$-type (20-25 GPa) and α-$PbO_2$-type structures are meta-stable in $IrO_2$, confirming previous experimental observation of rutile-type transforming directly into pyrite-type structure ($Pa\overline{3}$) [35]. This is in contrast with our study, documenting rather the $CaCl_2$-type structure and its development at pressures above 14 GPa. Following the evolution of the structure up to 20 GPa, a structural transformation sequence: rutile – $CaCl_2$ – α-$PbO_2$ - pyrite could be realised in $IrO_2$ at room temperature, considering that other phases are kinetically hindered. Experiment up to pressures higher than 20 GPa is needed to confirm this scenario.

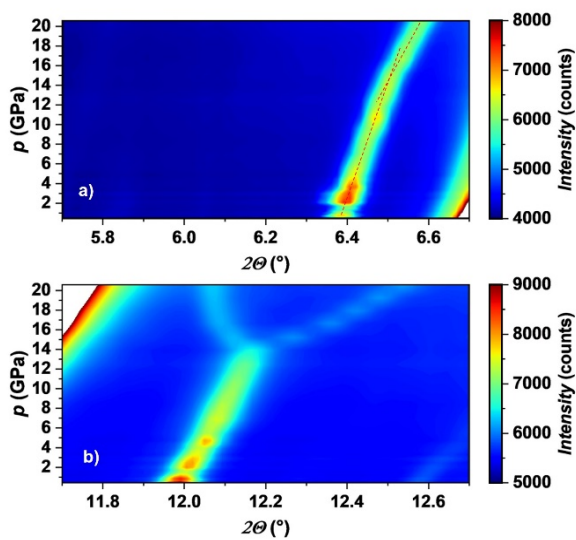

Fig. 2: Pressure evolution of selected peaks of the $IrO_2$ minority phase in the $Tm_2Ir_2O_7$ sample. a) At 15 GPa, a change of slope in the position evolution of the 110 peak is observed (dashed red lines are guides to the eye). b) The splitting of the low-pressure 211 peak is also observed around 15 GPa, indicating a lattice distortion. The splitting becomes more pronounced with further pressure application. The white regions in the corners of the figures are parts of out-of-scale pyrochlore peaks with a significantly higher intensity.

### 4. Temperature evolution of the $A_2Ir_2O_7$ structure; low-temperature powder diffraction:

The crystal structure of the investigated members (A = Pr, Sm, Dy-Ho), sufficiently characterising the $A_2Ir_2O_7$ series, was identified with the ordered cubic structure of the pyrochlore type (space group $Fd\bar{3}m$) in the whole temperature range. See Fig. 3a with the diffraction pattern of $Lu_2Ir_2O_7$ fitted with the pyrochlore model as an example; data and fits of the remaining members are presented in the supplementary material Fig. S1 and S2. A very good agreement between the data and the fit is presented by zooming in on the high-angle region of the data (inset in Fig. 3a) or following the difference curve. In addition to pyrochlore reflections, weak diffraction peaks are observed in the majority of the samples, which could be unambiguously identified and fitted with $A_2O_3$ (space group $Ia\text{-}3$ for A = Sm, Dy-Lu [42] and space group $P\bar{3}m1$ for A = Pr [43]), $IrO_2$ (space group $P4_2/mnm$ [35,36]), and Ir (face-centered cubic) minority phases. We highlight sharp Ir peaks on top of the broader pyrochlore peaks, which are well visible in diffraction patterns (inset of Fig. 3a), despite the fact that the determined weight percentage of Ir in the sample is well below 1%. Unlike $IrO_2$ (see previous section), the Ir minority phase remained stable in the whole investigated temperature and pressure range.

The diffraction peaks described by the pyrochlore-type structure are relatively broad and asymmetric, considering the energy and angle resolution of the instruments and the temperature of measurement of 4 K. The asymmetry of the pyrochlore peaks is demonstrated on the 440 reflection; see the weakly asymmetric peak of $Lu_2Ir_2O_7$ (Fig. 3b) and the significant asymmetry of the $Sm_2Ir_2O_7$ peak (Fig. 3c). We explain the asymmetry by assuming the presence of small fractions of off-stoichiometric (Ir-deficient) phases resulting in a distribution of lattice parameters in our samples. As our study focuses on both the temperature evolution of the lattice parameter $a$ and the free structural parameter $x_{48f}$ (strongly dependent on slight variations in diffraction intensity), it is imperative to describe the peak shape correctly to extract accurate integrated scattering intensities. Therefore, multiple pyrochlore phases (up to 11) with linearly varied lattice parameter $a$ and Ir/A occupancy were introduced refining the data. A similar method was previously used for off-stoichiometric structural refinements of, e.g., rare-earth $A_2Zr_2O_7$ zirconates [44] or Li-N-H structures [45]. The occupancy of the off-stoichiometric phases and the corresponding lattice parameters were evaluated using Vegard's law: stoichiometric $A_2Ir_2O_7$ represented one end of the series as the 100% Ir analogue and the $A_2O_3$ oxide was used as the 0% Ir analogue. Vegard's law gives only an approximate lattice parameter change with off-stoichiometry; also due to $A_2O_3$ crystallizing in a cubic structure (space group $Ia\text{-}3$) different than the pyrochlore one. To avoid overfitting, all pyrochlore phases shared the same shape parameters, thermal parameters, and $x_{48f}$ parameter, leaving only individual scale factors free. Fitting the diffraction data with 11 phases covering the (off-)stoichiometry region of 70-100% Ir led to the following results: (i) majority of the peak intensity is assigned to the 100% Ir phase (far right peak and corresponding Bragg reflection tick in Fig. 3b and Fig. 3c); (ii) smaller but still significant contribution to the peak intensity is connected with the 91-97% Ir phases; and (iii) phases with lower Ir content contribute only slightly or negligibly (see Table S2 and Fig. S3). The lattice parameter of the 100% Ir phase was used for further compressibility analysis, best representing the stoichiometric $A_2Ir_2O_7$. The ratio of the weight percentages was fixed based on the refinement of the low-temperature data. Subsequently, sequential fitting of diffraction patterns measured at higher temperature was performed using the fixed weight percentages of the refined phases.

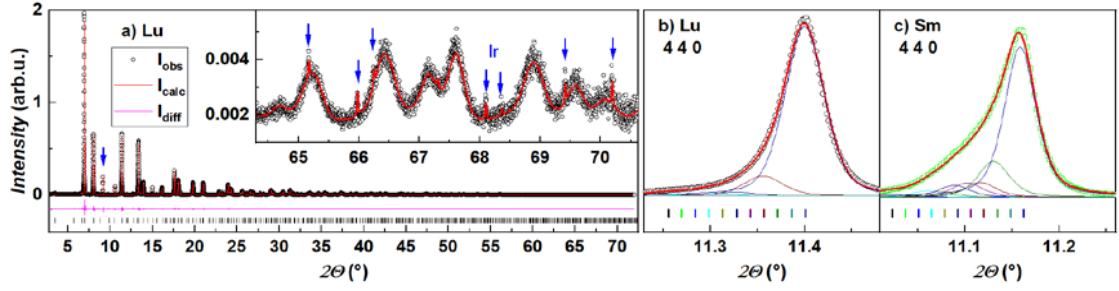

Fig. 3: Powder diffraction patterns of $A_2Ir_2O_7$ members collected at the ID22 beamline (ESRF, Grenoble) at 4 K. a) $Lu_2Ir_2O_7$ data are presented together with their high-angle inset, showcasing the instrument sensitivity and resolution despite the intensity being 500 times lower at high angles compared to low angles. In addition to the broader pyrochlore peaks, small sharp peaks (marked by blue arrows) are observed and identified as the pure Ir minority phase (with less than 1% mass of the sample). Ticks representing the minority phases and the off-stoichiometric pyrochlore phases are not shown for clarity (see Fig. S1 and S2 in the supplementary). Panels b) and c) show the refined pyrochlore 440 peak of $Lu_2Ir_2O_7$ and $Sm_2Ir_2O_7$, respectively, displaying also individual off-stoichiometric phases (respected ticks included). The low-asymmetry peak of $Lu_2Ir_2O_7$ is sufficiently described already by two phases, while more off-stoichiometric phases must be considered to fully describe the respective peak of $Sm_2Ir_2O_7$.

The structural parameters of the investigated $A_2Ir_2O_7$ members were refined using the same model at all temperatures, starting at 4 or 25 K, depending on the instrument, and increasing temperature up to 295 K. No structural transition was found throughout the experiments; all members retained the pyrochlore structure. Focusing first on the lattice parameter of the cubic lattice $a$, its temperature evolution was used to calculate the thermal expansivity shown in Fig. 4. The smooth evolution of the volume with temperature is observed. No indications of phase transition or clear anomalies connected with the magnetic ordering around $T_{Ir}$ (arrows in Fig. 4) were evidenced. We note that the jump at 50 K in $Sm_2Ir_2O_7$ and $Lu_2Ir_2O_7$ data is purely instrumental in origin. Data below 50 K used a beam attenuated to 50% intensity to reduce the beam-induced heating. Despite the expectation that this would be minimal above 50 K and compensated by cryostat cooling, an anomaly in the lattice thermal expansivity is still observed. Unfortunately, the magnitude of the sample heating with and without the attenuator cannot be reasonably and quantitatively determined. The $Pr_2Ir_2O_7$ data reveal, in addition to the jump at 50 K, some additional scatter that we ascribe to the instrument and fitting error. In turn, the ID22 data cannot be analysed supportably; only a very tentative fit of $Pr_2Ir_2O_7$ expansivity was performed. The remaining members, measured using the KMC-2 beamline, with significantly lower photon flux, show smooth dependencies of the expansivity. The expansivity was fitted using a combination of Debye's model and Grüneisen theory [46,47]:

$$V = V_0 + I_V T F(\theta_D/T), \quad (1)$$

where $V_0$ is the zero-temperature volume, $I_V = 3k_B r\gamma/K_0$, and $F(\theta_D/T)$ is the analytical Debye integral. Three parameters $V_0$, $I_V$, and $\theta_D$ are refined in the model. Thacher's approximation [46] was used on the Debye integral to model both low- and high-temperature regions numerically. The refined $I_V$ parameter can be used to estimate the Grüneisen parameter $\gamma$ in the case that the isothermal bulk modulus $K_0$ is known ($\gamma = I_V K_0/3k_B r$ with $r = 88$ atoms per unit cell for $A_2Ir_2O_7$ iridates). The refined parameters are listed in Table 1.

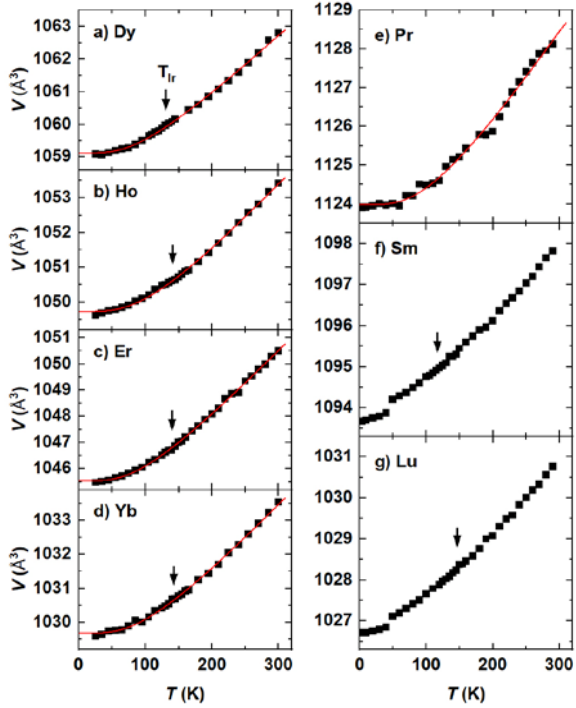

Fig. 4: Thermal expansivity of $A_2Ir_2O_7$ iridates measured at the KMC-2 (A = Dy, Ho, Er, Yb) and the ID22 (A = Pr, Sm, Lu) beamlines. The data were fit to the Debye model (1) (red lines). No anomaly is observed around $T_{Ir}$ (indicated by the arrows). The jump in expansivity at 50 K in A = Pr, Sm and Lu members is an instrumental artefact connected with using the beam attenuator at low temperatures; see the text for explanation.

The second free structural parameter $x_{48f}$ dictates the distortion of the octahedral oxygen cages around the Ir cations in the pyrochlore lattice [48]. Ir-O-Ir bond angle is purely dependent on the $x_{48f}$ parameter, while the Ir-O bond length is also dependent on the lattice parameter $a$. Fig. 5 shows the temperature evolution of $x_{48f}$ and Ir-O bond length for the $A_2Ir_2O_7$ members measured at the ID22 beamline. An increase in the $x_{48f}$ parameter with the increasing atomic number of A is followed. In general, a monotonous decrease of $x_{48f}$ is observed with increasing temperature in all the presented data. We did not reproduce the results of the recent powder diffraction study on $Eu_2Ir_2O_7$ done by Das et al. [25]. The previous data, presented without any associated error bars, revealed an anomaly in $x_{48f}$ (and bond angle and bond length, see Fig. 5c) at the temperature of the phase transition $T_{Ir}$; that is, the temperature of magnetic ordering and metal-insulator transition. With the exception of $Pr_2Ir_2O_7$, all other investigated $A_2Ir_2O_7$ members reveal the magnetic order and transition to the insulating state in magnetization and electrical resistivity data (we investigated the identical samples in our previous works [11,49,50] and this work). Therefore, an anomaly is indeed expected not only for $Eu_2Ir_2O_7$, but also for the other $A_2Ir_2O_7$ members. Despite the high resolution of the powder diffraction data and a very good agreement with the structural model (Fig. 3a), we were unable to confirm any anomaly in the A = Sm and Lu members. No anomaly on respective dependencies in Fig. 5 is followed within our experimental error. $Pr_2Ir_2O_7$ remains paramagnetic down to the lowest temperature [15]. Therefore, no anomaly in the data is expected or has been observed. The data measured using the KMC-2 beamline, A = Dy, Ho, Er, and Yb members, are connected with even larger error bars and do not allow for a similar analysis.

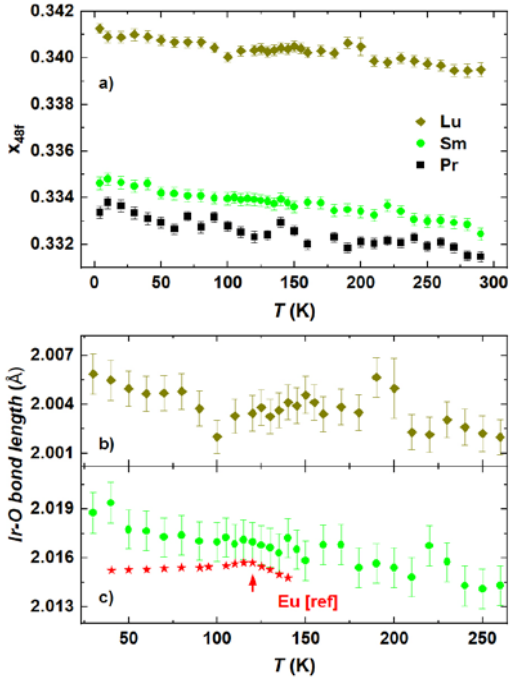

Fig. 5: Temperature evolution of a) the fractional coordinate $x_{48f}$ of the oxygen Wyckoff position 48f in $A_2Ir_2O_7$ iridates with A = Pr, Sm, and Lu; and b) and c) the Ir-O bond length (zoomed-in temperature range) in $Lu_2Ir_2O_7$ and $Sm_2Ir_2O_7$, respectively. No clear anomaly in the data is observed; just a decreasing tendency of both parameters with increasing temperature is followed. For comparison, the data of $Eu_2Ir_2O_7$ from Das et al. [25] is also plotted (error bars not reported), showing an anomaly close to the transition temperature $T_{Ir}$ (red arrow).

## 5. $A_2Ir_2O_7$ pyrochlore lattice under pressure:

Five $A_2Ir_2O_7$ members (A = Pr, Sm, Ho, Tm, and Lu), were investigated under applied pressure at ambient temperature. Synchrotron X-ray diffraction patterns of $Pr_2Ir_2O_7$ were measured using the ID15b beamline of the ESRF, the other members were studied employing the I15 beamline of the Diamond light source. The pyrochlore structure of investigated iridates was shown to be robust against applied pressure up to 20 GPa. No structural transition is seen in the diffraction patterns, as evidenced on the contour map in Fig. 6a. Traces of the minority phases were observed on top of the background, in good agreement with the higher-resolution measurements without the DAC, discussed above. Coincidentally, a previously unreported structural phase transition of the $IrO_2$ impurity phase in applied pressure was followed in the data; see section 3. An example of the diffraction patterns of $Lu_2Ir_2O_7$ is presented in Fig. 6b-d. The data of remaining members are presented in the supplementary material Fig. S4. An excellent agreement between the data and the structure model (Rietveld analysis) is observed. Similarly as in the case of the above-discussed data, an asymmetry of the diffraction peaks complicated the fitting process. Nevertheless, the experimental peak broadening in applied pressure, to a great extent, masks the asymmetry. As a result, only 2 or 3 pyrochlore phases were used to describe the peak shape. The fitting process was the same as described in the previous section. The ratio of the weight percentages was refined at ambient pressure and fixed for sequential pressure-dependent refinements. The insets in Fig. 6c-d shows a comparison between the peak shapes in ambient and high pressures. Although helium used as the pressure medium solidifies above approximately 11.5 GPa at ambient temperature, leading to quasi-hydrostatic conditions, the high-pressure experimental data are not visibly influenced.

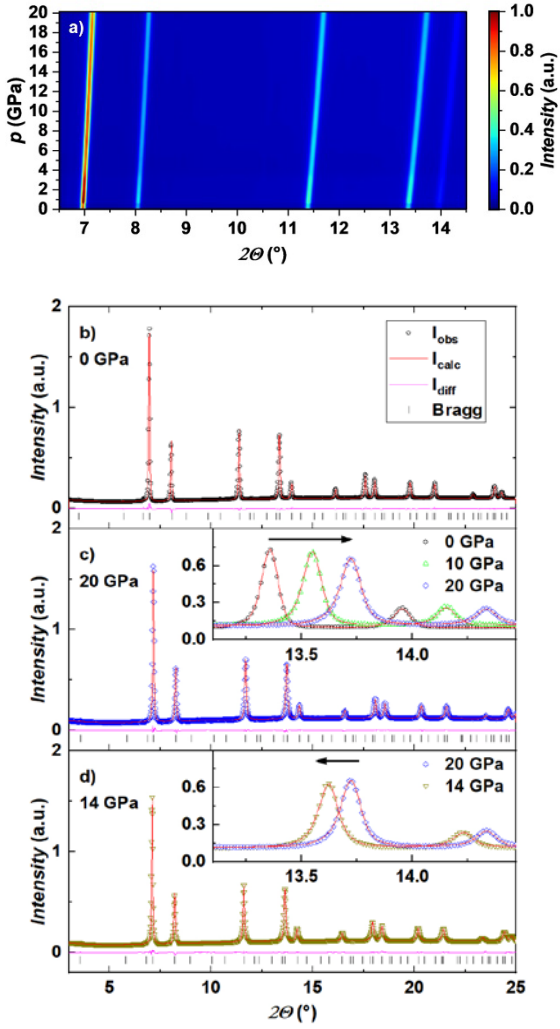

Fig. 6: High-pressure diffraction patterns of $Lu_2Ir_2O_7$ measured at the I15 beamline. a) Pressure evolution of the low-angle pyrochlore peaks' positions. No structural change is observed at pressures up to 20 GPa. Refined diffraction pattern measured in the DAC b) at ambient pressure, c) at the highest measured pressure of 20 GPa, and d) after depressurization of the DAC membrane, with a residual pressure of 14 GPa. The insets show the pressure evolution of the peak shape; only relatively small broadening was observed even at 20 GPa. Arrows indicate the pressure history. Ticks representing the minority phases and the off-stoichiometric pyrochlore phases are not shown for clarity (See Fig. S4 in the supplementary).

Pressure evolution of the lattice parameter $a$ is characterized by the volume compressibility in Fig. 7. Third-order Birch-Murnaghan equation of state [34]

$$P = \frac{3K}{2}\left[\left(\frac{V_0}{V}\right)^{\frac{7}{3}} - \left(\frac{V_0}{V}\right)^{\frac{5}{3}}\right]\left[1 + \frac{3}{4}(K' - 4)\left(\left(\frac{V_0}{V}\right)^{\frac{2}{3}} - 1\right)\right] \quad (2)$$

was used to model the compressibility data. $K$ denotes the bulk modulus and $K'$ its first derivative. The Murnaghan and Vinet equations of state [34] produced nearly identical results, which is not surprising considering that the total compression at 20 GPa was not higher than approximately 10% of the lattice volume. The results of the respective fits to the compressibility data are listed in Table 1. In the case of A = Pr and Ho members, the fitted bulk modulus was used to estimate the Grüneisen parameter $\gamma$ based on the refined temperature-dependent properties discussed above.

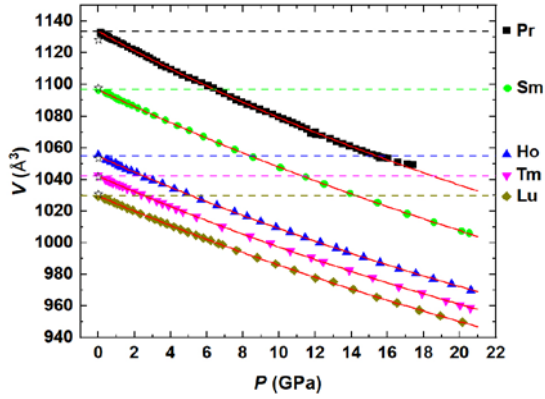

Fig. 7: High-pressure compressibility of $A_2Ir_2O_7$ iridates (A = Pr, Sm, Ho, Tm, and Lu) at room temperature. Solid red lines are fits to the Birch-Murnaghan equation of state (2). Dashed coloured lines show the ambient volumes of the respective iridates to guide the eye. Empty stars at 0 GPa indicate the volume determined by high-resolution room-temperature diffraction at ambient pressure (Section 4.); good agreement with the DAC data is demonstrated.

In contrast to the pronounced evolution of the lattice parameter/volume with applied pressure, the $x_{48f}$ parameter remains virtually constant up to 20 GPa (see Fig. 8). Compared to the ambient pressure experiments on relatively large samples, the high-pressure diffraction experiments are generally less accurate, which is reflected in the large error bars connected with the refined $x_{48f}$ parameter. Therefore, only pronounced trends in pressure evolution can be followed. The $x_{48f}$ parameter generally increases with increasing atomic number of A, confirming the ambient pressure results presented in Section 4. However, the $Tm_2Ir_2O_7$ data with the largest error bars (larger amount of minority phases than the rest of the studied members) breaks this tendency. The pressure evolution of $x_{48f}$ reveals a constant behaviour, as any potential smaller-scale tendencies are overshadowed by the experimental error. The local minimum observed in the $Lu_2Ir_2O_7$ data at pressures around 2 GPa is attributed to instrumental and refinement artefacts.

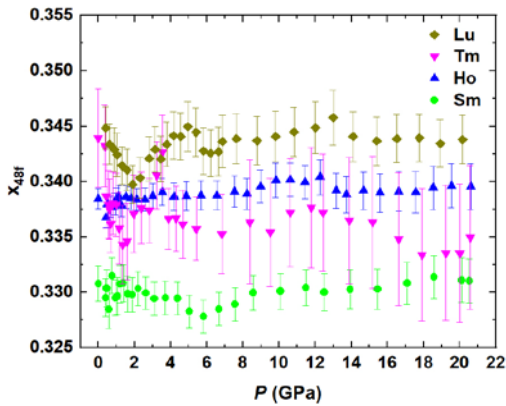

Fig. 8: Pressure evolution of the fractional coordinate $x_{48f}$ of the oxygen Wyckoff site 48f in $A_2Ir_2O_7$ iridates with A = Sm, Ho, Tm, and Lu. Large error of determined parameters does not allow to reliably discuss any trends observed in the data.

Table 1: Refined structural and compressibility parameters of $A_2Ir_2O_7$ iridates and their statistical errors. $a_T$ and $a_P$ are lattice parameters in ambient conditions from temperature- and pressure-dependent diffraction experiments (corresponding to the 100% Ir phase), respectively. Zero-temperature volume estimation $V_0$, Debye temperature estimation $\Theta_D$, isothermal bulk modulus $K$ and its derivative $K'$, thermal parameter $I_V$, and Grüneisen parameter estimation $\gamma$ are presented.

| A | $a_T$ (Å) | $a_P$ (Å) | $V_0$ (Å$^3$) | $\Theta_D$ (K) | $K$ (GPa) | $K'$ | $I_V$ (10$^{-2}$Å$^3$/K) | $\gamma$ |
|---|---|---|---|---|---|---|---|---|
| Pr | 10.41002(2) | 10.4242(1) | 1123.97(5) | 420(35) | 181.2(17) | 4.7(3) | 2.62(14) | 1.3(1) |
| Sm | 10.31597(2) | 10.3116(2) | - | - | 200.7(9) | 3.81(11) | - | - |
| Dy | 10.20512(19) | - | 1059.12(3) | 359(20) | - | - | 1.93(5) | - |
| Ho | 10.17499(13) | 10.1811(1) | 1049.73(2) | 407(16) | 202.6(15) | 4.7(2) | 2.11(5) | 1.2(1) |
| Er | 10.16556(23) | - | 1045.53(2) | 365(11) | - | - | 2.70(4) | - |
| Tm | - | 10.1396(3) | - | - | 206.2(11) | 4.45(16) | - | - |
| Yb | 10.11056(19) | - | 1029.68(2) | 386(16) | - | - | 2.11(5) | - |
| Lu | 10.10154(2) | 10.0968(1) | - | - | 210.4(11) | 4.23(14) | - | - |

### 6. Stability of pyrochlore structure in $A_2Ir_2O_7$ upon temperature, pressure and chemical pressure application:

Inspecting the diffraction data and evolutions of refined structural parameters, several observations are discussed. First, both the lattice parameter $a$ and the fractional coordinate $x_{48f}$ develop smoothly with rare-earth cation substitution (chemical pressure) in $A_2Ir_2O_7$. The lattice parameter decreases with decreasing A size; the standard lanthanide contraction is observed (Table 1). While the pyrochlore lattice is contracted with increasing atomic number of A, the parameter $x_{48f}$ increases. Such a conclusion can be made considering the investigated samples were synthesized using the same preparation method, studied using the same protocols at high-resolution X-ray synchrotron diffractometers and refined by the same software. Comparing the refined $x_{48f}$ parameters with those of independent studies [5,11,51,52,53,54,55,56,57], a qualitative agreement and a similar increasing tendency with atomic number is observed.

The trend of parameter $x_{48f}$ increasing with decreasing lattice volume is further demonstrated following the low-temperature data in Fig. 4 and Fig. 5. The lattice parameter decreases with decreasing temperature, while $x_{48f}$ slightly increases. The same development was also observed in the paramagnetic state of $Eu_2Ir_2O_7$ [25]. As the $x_{48f}$ parameter dictates the distortion of the octahedral oxygen cage around the Ir cations, a larger distortion is manifested when moving away from the ideal $x_{48f} = 0.3125$ case. The largest distortion in the series is revealed in $Lu_2Ir_2O_7$, especially when reaching the lowest temperatures. One can speculate that a further decrease of the lattice volume, induced by, e.g., substitution of Ir, would lead to the value of 0.375. The local environment would resemble the defect-fluorite case [58,59], but the cations and oxygen vacancies would remain ordered. Increase of the distortion can also be tied to the magnetic properties, that is, the magnetic ordering temperature $T_{Ir}$. We propose two competing mechanisms for $T_{Ir}$ suppression in the $A_2Ir_2O_7$ series, depending primarily on the the local $IrO_6$ octahedral environment. The magnetic ground-state, and $T_{Ir}$, is determined by both the electronic correlation strength, $U/t$, tuned by adjusting the bandwith primarily by isotropic lattice parameter adjustment, and the strength of the $Ir^{4+}$ Ising spin-anisotropy, brought about by the magnitude of the tetragonal distortion of the $IrO_6$ octahedron. Pressure drives the system to a (non-magnetic) metallic state through increased bandwidth; while chemical substitution, although also decreasing the volume, leads to an increased trigonal distortion which strengthens the Ising spin-anisotropy and promotes the magnetically ordered insulating state. Indeed, $T_{Ir}$ generally increases with the atomic number of A, which can also be said about the distortion dictated by $x_{48f}$. However, significant shift in $T_{Ir}$ that can be seen between A = Pr and Sm (compared to the slight shift between A = Sm to Lu) is not reproduced in the $x_{48f}$ evolution, which shows only slight increase throughout the whole series instead (see Fig. S6).

Investigating the pressure compressibility of the $A_2Ir_2O_7$ iridates with A = Pr, Sm, Ho, Tm, and Lu enabled us to cover, to a large extent, the whole series. Interpolating structural parameters allows a reasonable estimation of properties for the rest of the members. Therefore, no structural transitions are expected for any of the $A_2Ir_2O_7$ members, at least up to 20 GPa (Fig. 7). Moreover, the compressibility of, e.g., in pressure uninvestigated A = Yb member can be estimated by combining compressibilities of the A = Tm and Lu members. The robustness of the pyrochlore structure in $A_2Ir_2O_7$ iridates makes this series suitable for studies of not only structural but also magnetic and conducting properties with respect to chemical and external pressure. That is, external pressure can be compared with chemical pressure based on the change in the lattice parameter. For example, an application of pressure on $Pr_2Ir_2O_7$ leads to a compression of the lattice; that is, decreasing the lattice parameter/volume across the rare-earth series, and reaching the lattice parameter of $Yb_2Ir_2O_7$ at around 20 GPa. Such a change in the volume should be connected with a considerable change of the $x_{48f}$ parameter. This is, however, not observed in the data (Fig. 8). No clear evolution of $x_{48f}$ with applied pressure is followed, considering the error of the parameter determination. Instead, the $x_{48f}$ remains practically constant up to 20 GPa. Similarly to our results, no (clear) evolution of the $x_{48f}$ parameter with external pressure was reported for $Eu_2Ir_2O_7$ in Ref. [60]. On the other hand, a small increase of $x_{48f}$ with applied pressure was observed in [52]; although small compared to the variation with chemical pressure. Based on these observations, we can conclude that the external pressure impacts the distortion of oxygen octahedra in $A_2Ir_2O_7$ negligibly.

Thermal expansivity data were fitted and refined parameters are listed in Table 1. The determined Debye temperature $\Theta_D$, although its determination from thermal expansivity is less straightforward and accurate than from specific heat, is in good agreement with previous results [12,50,51,47]. In general, $\Theta_D$ values tend to fall into the approximate region of 350-400 K; slightly larger values for A = Pr and Ho members were revealed by our study. Similarly to the previous results [12,50,51,47], no clear dependence of Debye temperature on A was followed. Bulk modulus of the studied $A_2Ir_2O_7$ members is comparable with the recent independent results on $Sm_2Ir_2O_7$ (215.6 GPa) [21] and $Eu_2Ir_2O_7$ (209.6 GPa) [52]. A systematic increase of $K$ with the atomic number of rare-earth is observed (Table 1). The increase of the compressibility is perfectly consistent with decreasing volume with A; that is, a lanthanide contraction. We note that the determined bulk modulus is similar to that of the pyrochlore $A_2Zr_2O_7$ zirconates [62], which tend to exhibit a structural transition well below 20 GPa [62]. Of course, the substitution of Ir by Zr has a significant impact on the lattice volume (with a smaller volume of zirconate lattice). The Grüneisen parameter $\gamma$ estimated from the combination of both thermal and pressure experiments is almost unchanged comparing A = Pr and Ho members. The values are in excellent agreement with the previous investigation of thermal expansivity of $Tm_2Ir_2O_7$ (similar values of $I_a$) [47], and with the Grüneisen parameter of individual phonon modes in Raman spectroscopy of $Eu_2Ir_2O_7$ [60]. They are also comparable, for example, to the typical $\gamma$ values of pure metals [61].

7. **Conclusions**

Structural properties of the rare-earth $A_2Ir_2O_7$ members were investigated down to 4 K and up to 20 GPa, demonstrating the robustness of their structure. As the series was adequately covered investigating iridates with A = Pr, Sm, Dy, Ho, Er, Tm, Yb, and Lu, the conclusion of a stable pyrochlore structure within the whole $A_2Ir_2O_7$ series was made. Thermal compressibility was refined using the Debye model, showing comparable Debye temperature values to the previously published ones determined from specific heat data. The free fractional coordinate $x_{48f}$ of the oxygen anion increased upon cooling; however, no sign of an anomaly was observed. Thus, our data did not reproduce the anomaly at the temperature of the iridium magnetic ordering transition reported previously for A = Eu member. Pressure compressibility data were refined employing the third-order Birch-Murnaghan equation of state, showing a systematic increase of the bulk modulus along the lanthanide contraction (chemical pressure). No evolution of $x_{48f}$ was observed with pressure application. The Grüneisen parameter was estimated for selected members using isothermal bulk modulus and thermal compressibility data. Comprehensive structural investigation of the $A_2Ir_2O_7$ series paved the way for a better understanding and reliable interpretation of the physical properties of individual members regarding temperature evolution and external and chemical pressure application. In addition, the crystal structure of the $IrO_2$ minority phase was investigated. The tetragonal rutile-type structure of $IrO_2$ was shown to be stable down to 4 K and up to 14 GPa. Upon further pressure application, the lattice was observed to orthorhombically distort. The lattice parameters of the orthorhombic structure at 20 GPa were determined and space group of the high-pressure phase structure suggested.

## 8. Acknowledgements:

We acknowledge the European Synchrotron Radiation Facility, Grenoble, France, for awarding us an experimental time at the ID22 beamline (experiment HC-5197; DOI: 10.15151/ESRF-ES-1178037339) and ID15b beamline (experiment HC-4856; DOI: 10.15151/ESRF-ES-761061310). We appreciate the Helmholtz-Zentrum Berlin, Germany, for awarding us an experimental time at the KMC-2 beamline (experiment 231-11842-ST). We thank also the Diamond Light Source, Didcot, UK, for awarding us an experimental time at the I15 beamline (experiment CY30424). We are grateful for the help of A. Fitch during our experiment at the ID22. Sample synthesis and characterization was done in MGML (mgml.eu), which is supported within the program of Czech Research Infrastructures (project no. LM2023065). The study was supported by the Charles University, project GA UK (no. 148622), and Barrande Mobility project (no. 8J24FR013).

**Supplementary material:**

Our extensive synchrotron X-ray diffraction studies of the $A_2Ir_2O_7$ compounds include a large amount of data and subsequent Rietveld refinement analysis. To maintain a reasonable length and clarity of the article, we present a substantial portion of the data within these supplementary materials.

The individual tables and figures are listed in the order they are referred to in the body of the article. No further explanatory text is included; see the body of the article for details.

Table S1: $IrO_2$ minority phase peaks identified in the diffraction patterns of the $Tm_2Ir_2O_7$ samples. The structure was identified as rutile-type at ambient pressure up to 13.9 GPa. At higher pressure, a splitting of some observed peaks was followed. At 20.6 GPa, nine out of thirteen identified peaks were split (in the main text, only eight most-intensive unambiguously determined peaks are discussed). Peak intensities could not be reasonably determined, due to the weak signal and vicinity to much more intensive majority phase pyrochlore peaks. The peaks determined with high ambiguity are denoted with *.

| P (GPa) | $2\vartheta$ Bragg positions of the reflections (deg.) | | | | | | | | | | |
|---|---|---|---|---|---|---|---|---|---|---|---|
| | 110 | 101 | | 200 | | 211 | | 220 | 310* | 112 | |
| 0 | 6.377 | 7.86 | | 9.035 | | 11.977 | | 12.781 | 14.299 | 14.399 | |
| 13.9 | 6.497 | 7.953 | | 9.211 | | 12.163 | | 13.032 | 14.536 | 14.584 | |
| 20.6 | 6.581 | 7.896 | 8.135 | 8.932 | 9.737 | 12.088 | 12.563 | 13.206 | 14.626 | 14.245 | 15.293 |
| | 301 | | 321* | | 222* | | 330* | | 312 | 411* | |
| 0 | 14.989 | | 17.569 | | 18.191 | | 19.241 | | 19.324 | 19.794 | |
| 13.9 | 15.27 | | 17.869 | | 18.536 | | 19.588 | | 19.595 | 20.104 | |
| 20.6 | 14.898 | 16.015 | 17.819 | 18.348 | 17.902 | 19.548 | 19.859 | 19.363 | 20.149 | 19.644 | 21.071 |

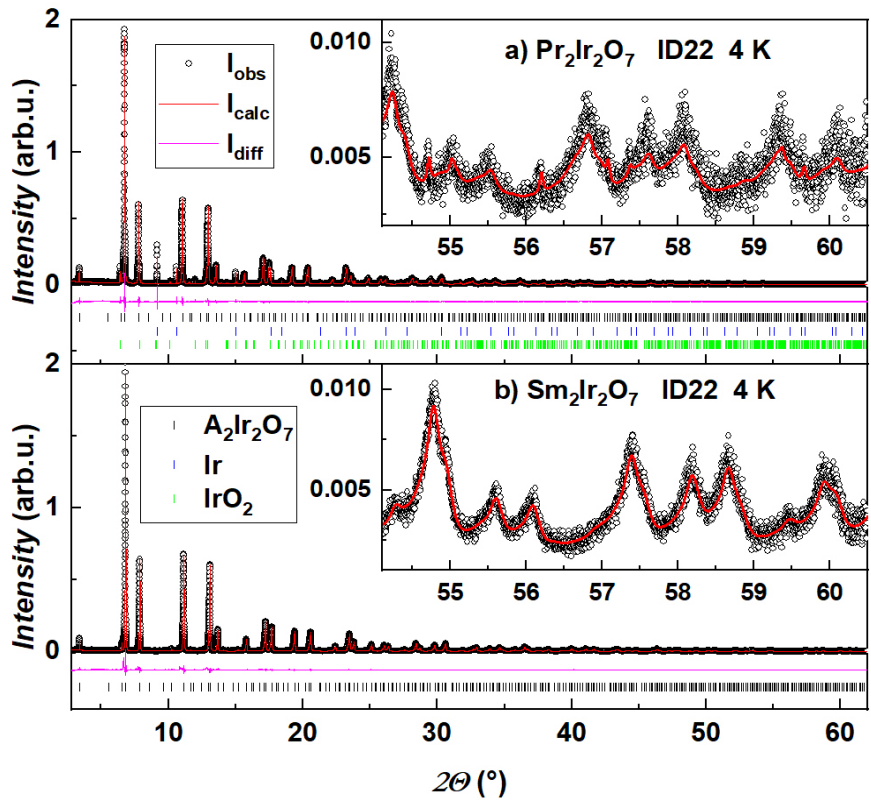

Fig. S1: Powder diffraction patterns of the a) $Pr_2Ir_2O_7$ and b) $Sm_2Ir_2O_7$ members collected at the ID22 beamline (ESRF, Grenoble) at 4 K. High-angle regions are presented in the insets. Just like in the case of the $Lu_2Ir_2O_7$ member (Fig. 3), also the $Pr_2Ir_2O_7$ pattern exhibits visible peaks from the pure Ir phase (approximately 1.4 wt.% of the sample; refined from the diffraction data), whereas no Ir peaks were detected in the $Sm_2Ir_2O_7$ member. Tick marks for the $A_2Ir_2O_7$ phase indicate the reflection positions of only the highest fraction (Ir-stoichiometric, and smallest lattice parameter) component.

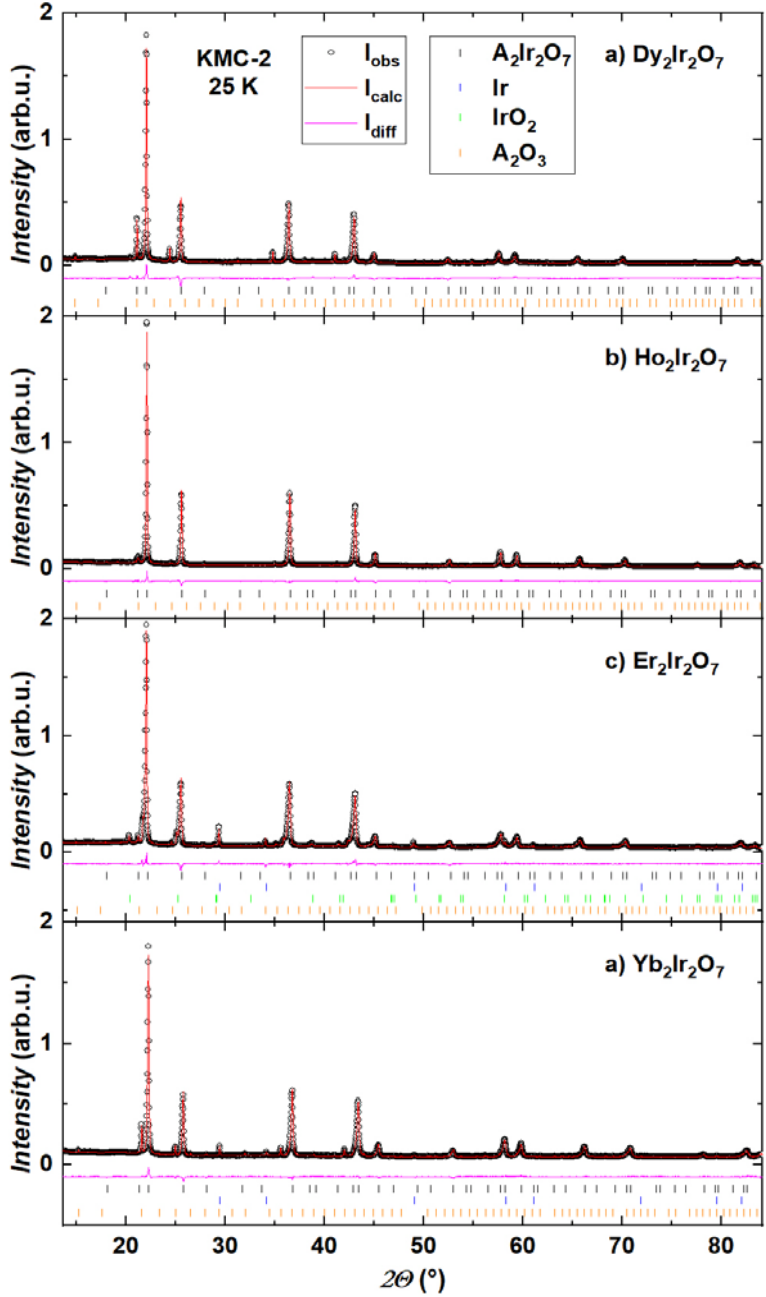

Fig. S2: Powder diffraction patterns of a) $Dy_2Ir_2O_7$, b) $Ho_2Ir_2O_7$, c) $Er_2Ir_2O_7$, and d) $Yb_2Ir_2O_7$ members collected at KMC-2 beamline (Bessy II, Berlin) at 25 K. Tick marks for the $A_2Ir_2O_7$ phase indicate the reflection positions of only the highest fraction (Ir-stoichiometric, and smallest lattice parameter) component.

Table S2: Analysis of the temperature-dependent experiments: weight percentages of 11 individual pyrochlore phases with different Ir occupancy. Every off-stoichiometric phase had the lattice parameter linearly shifted (by 0.013 Å) with occupancy. That is, lattice parameter $a_{100\%}$ is smaller than $a_{70\%}$ by 0.13 Å. $Er_2Ir_2O_7$ was the only exception, as a total shift of 0.16 Å was needed to model the peak shape correctly. The lattice parameter $a_{average}$ was calculated as a mean average of lattice parameters of the 11 individual phases with corresponding weight percentages/scale factors, resulting in slightly higher values compared to $a_{100\%}$.

| A | $a_{100\%}$ | $a_{average}$ | weight percentage of phases with different Ir occupancy | | | | | | | | | | |
|---|---|---|---|---|---|---|---|---|---|---|---|---|---|
| | | | 100% | 97% | 94% | 91% | 88% | 85% | 82% | 79% | 76% | 73% | 70% |
| Pr | 10.41002(2) | 10.43039(2) | 47.32 | 7.60 | 17.91 | 12.38 | 7.52 | 3.30 | 1.42 | 1.19 | 0.45 | 0.00 | 0.90 |
| Sm | 10.31597(2) | 10.33481(2) | 61.59 | 0.12 | 14.71 | 6.04 | 5.88 | 4.91 | 2.31 | 2.69 | 0.38 | 0.00 | 1.37 |
| Dy | 10.20512(19) | 10.21754(19) | 60.41 | 16.67 | 10.09 | 6.42 | 1.40 | 2.31 | 0.07 | 1.32 | 0.00 | 0.00 | 1.30 |
| Ho | 10.17499(13) | 10.18683(13) | 53.53 | 27.97 | 8.35 | 3.02 | 3.92 | 0.52 | 1.49 | 0.09 | 0.59 | 0.00 | 0.52 |
| Er | 10.16556(23) | 10.19275(23) | 65.27 | 0.00 | 2.37 | 17.29 | 1.06 | 2.35 | 2.08 | 2.86 | 0.00 | 0.00 | 6.72 |
| Yb | 10.11056(19) | 10.12655(19) | 56.20 | 14.56 | 10.38 | 7.36 | 2.84 | 4.53 | 0.00 | 2.63 | 0.41 | 0.00 | 1.09 |
| Lu | 10.10154(2) | 10.10445(2) | 85.62 | 0.02 | 0.00 | 9.94 | 0.00 | 1.86 | 0.00 | 1.58 | 0.00 | 0.00 | 0.97 |

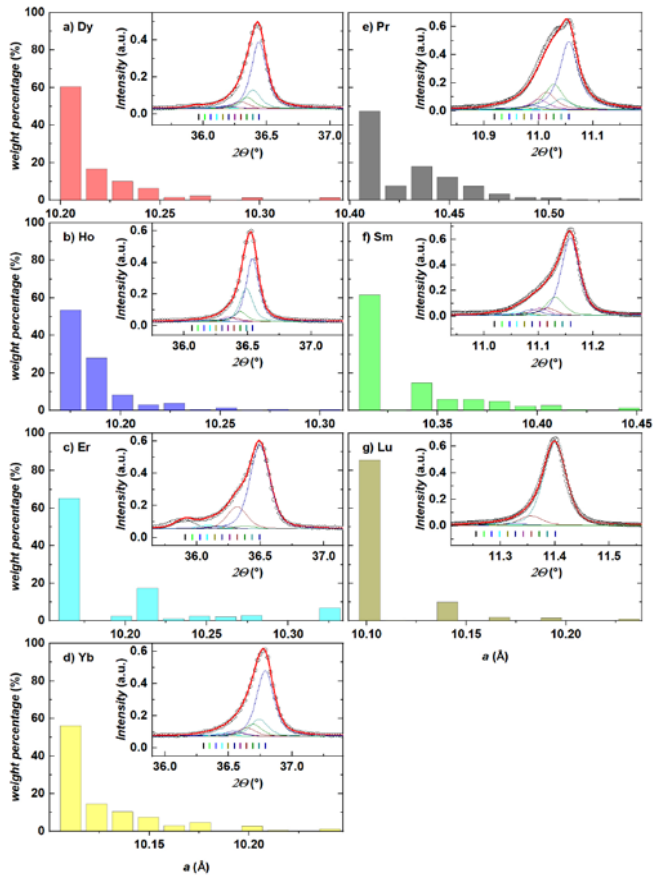

Fig S3: Refined weight percentages of the (off-)stoichiometric pyrochlore phases from thermal expansivity measurements: 4 K at ID22 (A = Pr, Sm, Lu) and 25 K at Berlin (A = Dy, Ho, Er, Yb). Corresponding peak shapes are shown in the insets. See Table S2 for more details.

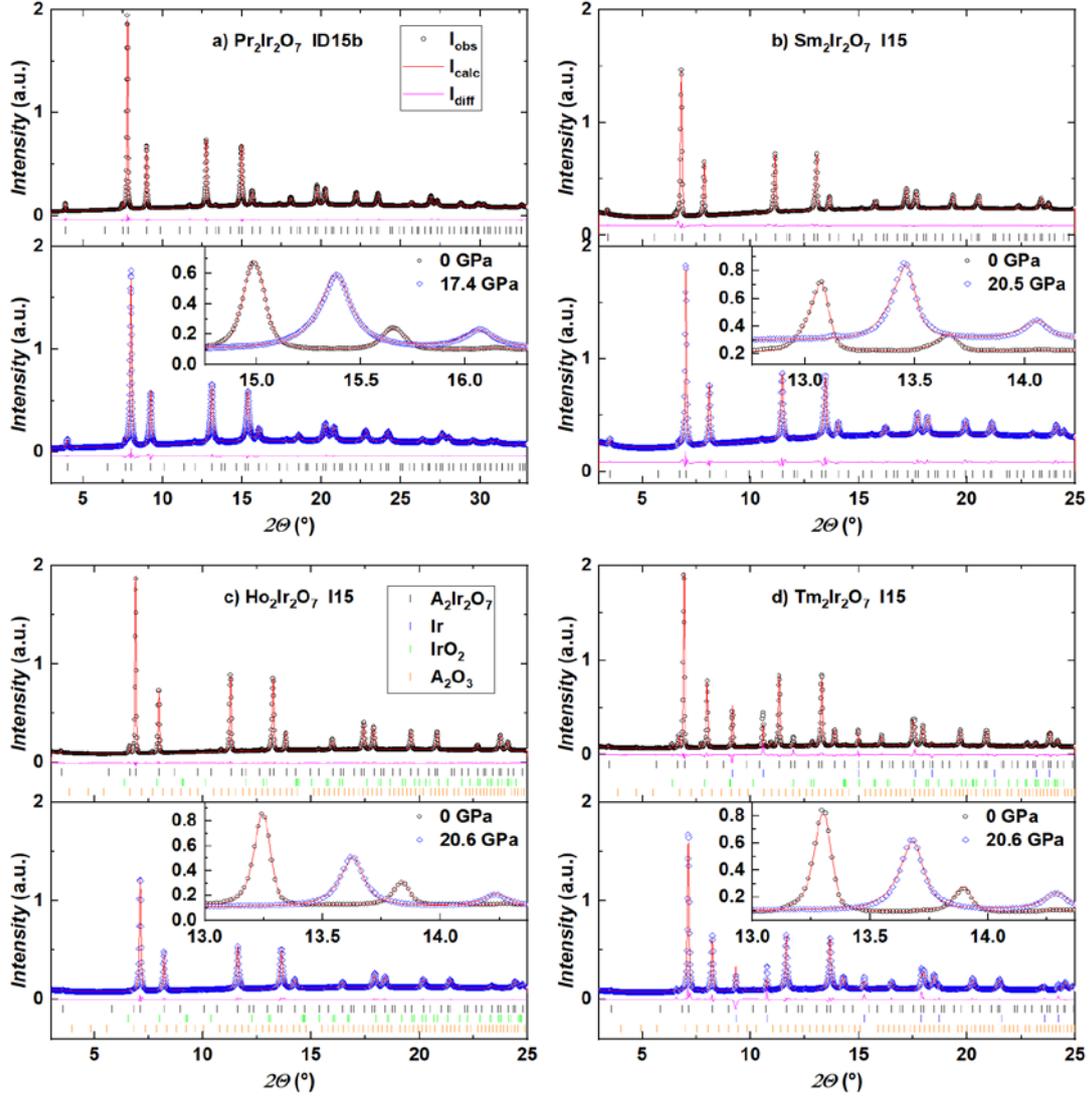

Fig. S4: High-pressure diffraction patterns of a) $Pr_2Ir_2O_7$ measured at ID15b and b) $Sm_2Ir_2O_7$, c) $Ho_2Ir_2O_7$, and d) $Tm_2Ir_2O_7$ measured at I15. The bottom panels show the data measured at the highest applied pressure. The insets show a zoomed-in evolution of the peak shape with applied pressure. Tick marks for the $A_2Ir_2O_7$ phase indicate the reflection positions of only the highest fraction (Ir-stoichiometric, and smallest lattice parameter) component.

Table S3: Analysis of the pressure-dependent experiments: weight percentages of 3 individual pyrochlore phases with different Ir occupancy. Similarly to ambient pressure experiments, every off-stoichiometric phase had the lattice parameter linearly shifted with occupancy, but in the case of pressure data the total shift to 80% was 0.09 Å.

| A | $a_{100\%}$ | $a_{average}$ | weight percentage of phases with different Ir occupancy | | |
| --- | --- | --- | --- | --- | --- |
| | | | 100% | 90% | 80% |
| **Pr** | 10.42425 | 10.42754 | 96.86 | 0.00  | 3.14 |
| **Sm** | 10.31157 | 10.32879 | 69.67 | 22.28 | 8.06 |
| **Ho** | 10.18111 | 10.188   | 87.23 | 10.06 | 2.70 |
| **Tm** | 10.1396  | 10.14836 | 84.70 | 10.95 | 4.35 |
| **Lu** | 10.09684 | 10.1022  | 90.89 | 6.13  | 2.98 |

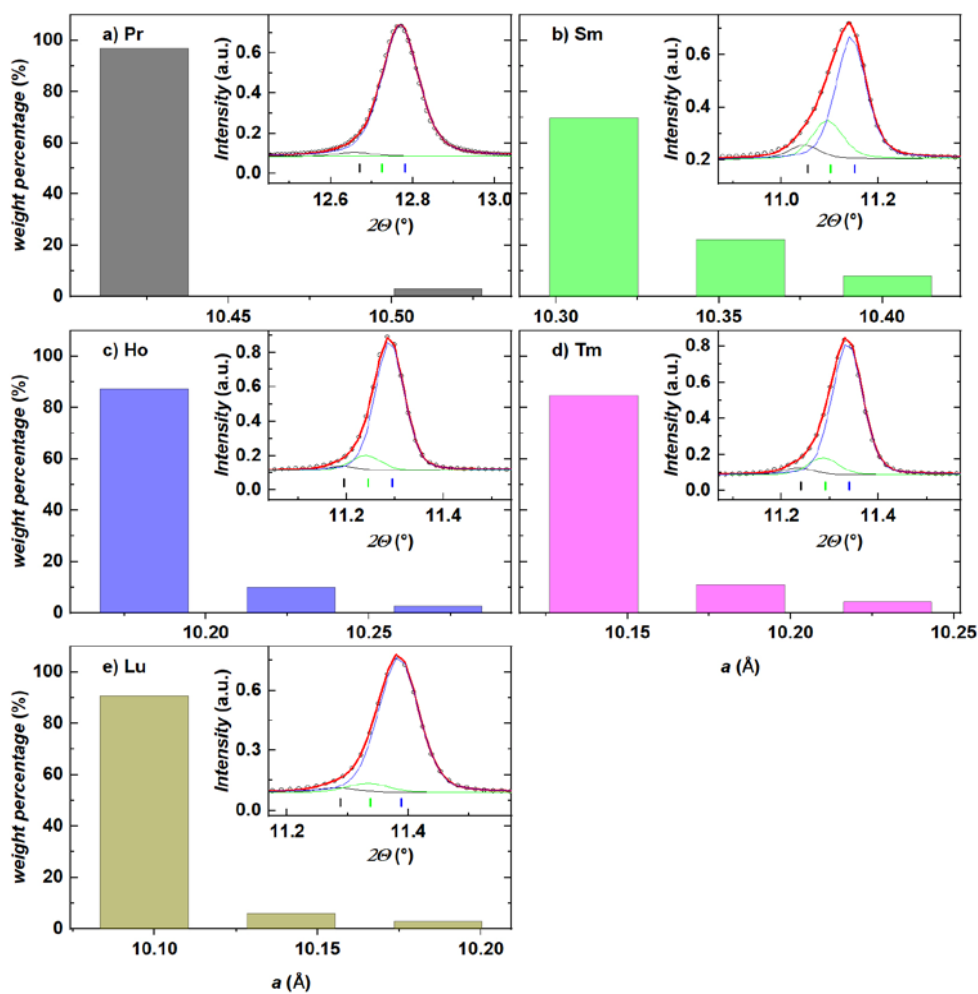

Fig S5: Refined weight percentages of the (off-)stoichiometric pyrochlore phases from pressure compressibility measurements. Corresponding peak shapes are shown in the insets. See Table S3 for more details.

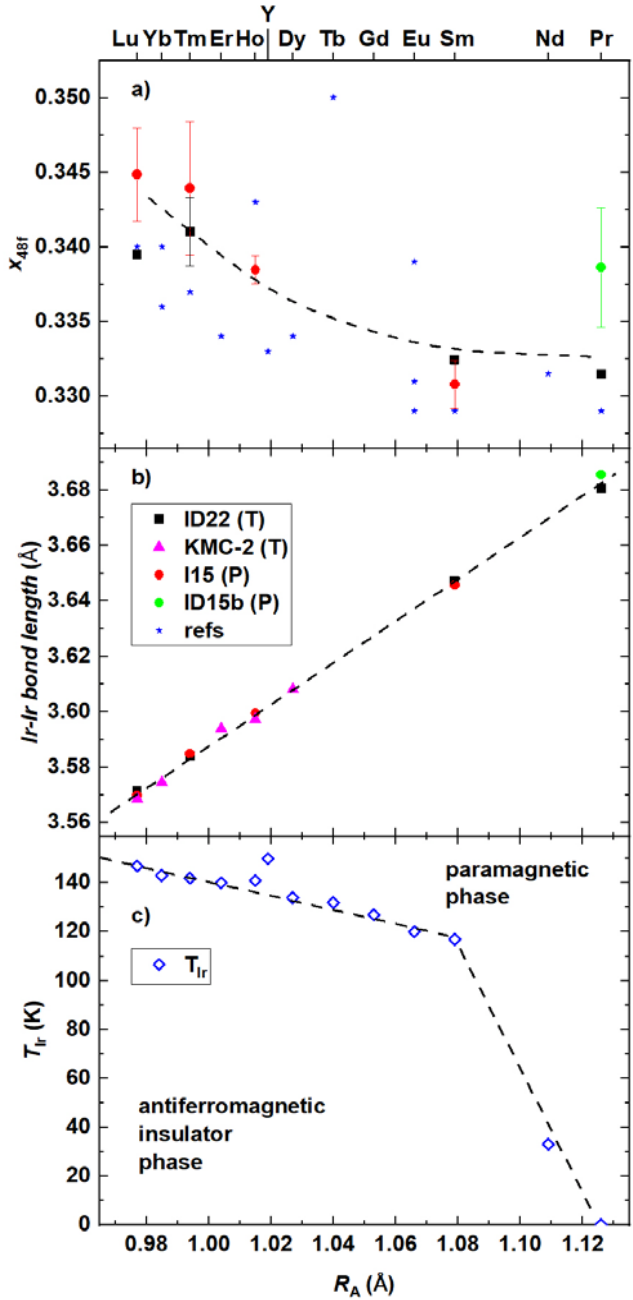

Fig S6: Rare-earth ionic radius dependence of structural (local) and magnetic properties of $A_2Ir_2O_7$ iridates. a) $x_{48f}$ fraction coordinate increases slightly with atomic number of A, while b) the Ir-Ir bond length (dependent on the lattice parameter) decreases linearly, following the lanthanide contraction. Local structural properties are compared to c) magnetic ordering temperature $T_{Ir}$ (of the Ir sublattice), which is concomitant with the metal-insulator (or semimetal-insulator) transition. In the region with the most drastic change of the magnetic properties (A = Sm to Pr), there is only negligible change in the $x_{48f}$ which dictates the distortion of the $IrO_6$ octahedra. Dashed lines are guides to the eye. Data of $T_{Ir}$ are adapted from [11] and data of $x_{48f}$ from [5,11,42,43,44,45,46,47,48].